\documentclass[aip,cha,amsmath,amssymb,reprint,author-year,author-numerical]{revtex4-1}
\usepackage{epsfig}
\usepackage{color}
\usepackage{graphicx}
\usepackage{dcolumn}
\usepackage{bm}
\input epsf
\usepackage[normalem]{ulem} % package used for "crossing out" text when doing comments on the paper

\graphicspath{{figures/}}

\begin{document}

%\preprint{AIP/123-QED}

\title{On controlling networks of limit-cycle oscillators}

\author{Per Sebastian Skardal}
\email{persebastian.skardal@trincoll.edu}
\affiliation{Department of Mathematics, Trinity College, Hartford, CT 06106, USA}

\author{Alex Arenas}
\affiliation{Departament d'Enginyeria Inform\`{a}tica i Matem\`{a}tiques, Universitat Rovira i Virgili, Tarragona, Spain}

\begin{abstract}
The control of network-coupled nonlinear dynamical systems is an active area of research in the nonlinear science community. Coupled oscillator networks represent a particularly important family of nonlinear systems, with applications ranging from the power grid to cardiac excitation. Here we study the control of network-coupled limit cycle oscillators, extending previous work that focused on phase oscillators. Based on stabilizing a target fixed point, our method aims to attain complete frequency synchronization, i.e., consensus, by applying control to as few oscillators as possible. We develop two types of control. The first type directs oscillators towards to larger amplitudes, while the second does not. We present numerical examples of both control types and comment on the potential failures of the method. 
\end{abstract}


\maketitle

\begin{quotation}
Collective rhythms in ensembles of interacting units generate novel phenomena in mathematics, physics, engineering, and biology~\cite{Strogatz2003,Pikovsky2003}. Moreover, robust collective rhythms characterized by synchronization is vital to the functionality of systems ranging from power grids~\cite{Motter2013NaturePhysics} and Josephson junction arrays~\cite{Wiesenfeld1996PRL} to cardiac tissue~\cite{Glass1988} and circadian rhythms~\cite{Yamaguchi2003Science}. This has motivated a need for control and optimization methods for coupled oscillator networks -- specifically towards attaining consensus among the individual oscillators~\cite{Dorfler2013PNAS,Skardal2014PRL}. In a recent publication we developed a simple control mechanism for attaining consensus in networks of coupled phase-oscillators based on identifying and stabilizing a target synchronized state~\cite{Skardal2015SciAdv}. Here we extend this this method to the case of networks of nonlinear limit-cycle oscillators, where the state of each oscillator is characterized not only by a phase angle, but also an amplitude~\cite{Matthews1990PRL}. While the presence of an amplitude for each oscillator yields a richer set of dynamical states overall, we find that consensus can still be attained in this more complicated scenario. 
\end{quotation}

\section{Introduction}\label{sec1}

Network-coupled dynamical systems are ubiquitous in nature and science~\cite{Strogatz2001Nature}, and as a result the control of such systems has been the focus of a great deal of research from the nonlinear dynamics and complex networks communities~\cite{Ott1990PRL,Motter2015Chaos}. For instance, the concept of controllability of complex networks has been established using control theory for linear dynamical systems~\cite{Lin1974IEEE,Liu2011Nature,Nepusz2012NatPhys,Yuan2013NatComm,Sun2013PRL}. Significant advances have also been made in the control of several nonlinear networks-connected dynamical systems~\cite{Grigoriev1997PRL,Wang2002PhysA,Li2004IEEE,Sahasrabudhe2011NatComm,Cornelius2013NatComm}. A particularly important family of network-coupled nonlinear dynamical systems that plays an important role in modeling phenomena ranging from synchronization of power grids~\cite{Rohden2012PRL} to cardiac excitation~\cite{Karma2013Rev} are networks of coupled oscillators~\cite{Arenas2008PR}. Control of coupled oscillator networks has also garnered significant attention recently~\cite{DeLellis2010IEEE,Wagemakers2014Chaos}. In a recent publication we presented a nonlinear dynamics-based method for controlling networks of coupled phase oscillators~\cite{Skardal2015SciAdv} -- specifically for attaining full frequency synchronization, i.e., consensus. In this paper we extend this control method to a more general and more complex family of coupled oscillator systems and explore its effectiveness in this more complicated scenario.

Limit-cycle oscillators hold an important place in the history of nonlinear science~\cite{Matthews1990PRL}. Here we consider a network of $N$ coupled Landau-Stuart oscillators $z_n$, $i=1\dots,N$, whose dynamics are governed by the following system of $N$ nonlinear, complex differential equations:
\begin{align}
\dot{z}_n=z_n(1-|z_n|^2+i\omega_n)+K\sum_{m=1}^NA_{nm}(z_m-z_n).\label{eq:LandauStuart}
\end{align}
In Eq.~(\ref{eq:LandauStuart}), $z_n\in\mathbb{C}$ describes the complex state of oscillator $n$, $\omega_n$ is the natural frequency of oscillator $n$, $K$ is the global coupling strength, and $[A_{nm}]$ is the adjacency matrix that encodes the network structure such that $A_{nm}=1$ if a link exists between nodes $n$ and $m$. (We consider here for simplicity the case of undirected networks such that $A^T=A$.) Interpreting the state of each oscillator $z_n$ as a phase $\theta_n$ and amplitude $\rho_n$, such that $z_n=\rho_ne^{i\theta_n}$, the dynamics of Eq.~(\ref{eq:LandauStuart}) can be written in terms of the evolutions of $\theta_n$ and $\rho_n$:
\begin{align}
\dot{\theta}_n&=\omega_n+\frac{K}{\rho_n}\sum_{m=1}^NA_{nm}\rho_m\sin(\theta_m-\theta_n),\label{eq:Polar01}\\
\dot{\rho}_n&=\rho_n\left(1-\rho_n^2\right)+K\sum_{m=1}^NA_{nm}\left[\rho_m\cos(\theta_m-\theta_n)-\rho_n\right],\label{eq:Polar02}
\end{align}
representing a natural generalization of the classical Kuramoto phase oscillator model~\cite{Kuramoto1984}.

Although some novel dynamical phenomena besides second-order phase transitions between incoherence and synchronization have recently been observed in simple Kuramoto phase oscillator networks~\cite{GomezGardenes2011PRL,Restrepo2014EPL,Skardal2015PRE}, the addition of an amplitude for each oscillator in Eq.~(\ref{eq:LandauStuart}) gives rise to a plethora of more robust dynamical phenomena, including amplitude death, period-doubling cascades enroute to chaos, extensive chaos, cluster states, and hysteresis~\cite{Mirollo1990JSP,Matthews1991PhysicaD,Takeuchi2013JPA,Ku2015Chaos}. Given the increase in complexity of the dynamics of limit-cycle oscillators in comparison to phase oscillators, a natural question arises in the control of limit-cycle oscillator networks. Can consensus still be reached in the case of limit cycle oscillator networks? Do the methods used to attain consensus in phase oscillator networks extend to the limit-cycle oscillator networks? In this paper, we address these questions.

Previously, control methods ranging from time-delay feedback~\cite{Choe2010PRE,Schneider2013Phil} and adaptive network structures~\cite{Lehnert2014PRE} have been applied to networks of limit-cycle oscillators. In this work we extend the control method presented in Ref.~\cite{Skardal2015SciAdv} to the limit-cycle oscillator dynamics given in Eq.~(\ref{eq:LandauStuart}). In fact, we present two distinct types of control. For type I control we prefer larger amplitudes, driving oscillators to the edge of the complex unit circle. For type II control we make no such preference, allowing for smaller amplitudes in the target state. We demonstrate the utility of both control types, and discuss the effect that each control type has on the macroscopic order parameters of the system.

The remainder of this paper is organized as follows. In Sec.~\ref{sec2} we present the control method. In Sec.~\ref{sec3} we present numerical examples of both control methods applied to random networks. In Sec.~\ref{sec4} we comment of some failures of the method. Finally, in Sec.~\ref{sec5} we conclude with a discussion of our results.

\section{Control Method}\label{sec2}

We begin by extending the control method present in Ref.~\cite{Skardal2015SciAdv} to the network-coupled Landau-Stuart model given in Eq.~(\ref{eq:LandauStuart}). We emphasize that our goal is to achieve a fully frequency-synchronized state characterized by $\dot{\theta}_1=\dots=\dot{\theta}_N$. We propose a simple linear feedback-type controller, adding to the right-hand-side of Eq.~(\ref{eq:LandauStuart}) a control term $f_n=F_n(z_n^*-z_n)$, obtaining
\begin{align}
\dot{z}_n=z_n(1-|z_n|^2+i\omega_n)&+K\sum_{m=1}^NA_{nm}(z_m-z_n)\nonumber\\
&\hskip4ex+F_n(z_n^*-z_n),\label{eq:Control}
\end{align}
where $F_n$ represents the control gain or strength applied to oscillator $n$ and $z_n^*=\rho_n^*e^{i\theta_n^*}$ is the target state for oscillator $n$. (We emphasize that $\cdot^*$ does not indicate complex conjugate, but rather the target value of a given quantity.) In particular, $F_n>0$ corresponds to some amount of control applied to oscillator $n$. The set of target states $z_n^*$ will be determined below. In polar representation, the addition of the control term results in the new set of equations:
\begin{align}
\dot{\theta}_n&=\omega_n+\frac{K}{\rho_n}\sum_{m=1}^NA_{nm}\rho_m\sin(\theta_m-\theta_n)\nonumber\\
&\hskip16ex+F_n\frac{\rho_n^*}{\rho_n}\sin(\theta_n^*-\theta_n),\label{eq:PolarControl01}\\
\dot{\rho}_n&=\rho_n\left(1-\rho_n^2\right)+K\sum_{m=1}^NA_{nm}\left[\rho_m\cos(\theta_m-\theta_n)-\rho_n\right]\nonumber\\
&\hskip16ex+F_n[\rho_n^*\cos(\theta_n^*-\theta_n)-\rho_n].\label{eq:PolarControl02}
\end{align}
We note that, for any finite combinations of frequencies $\omega_n$, coupling strength $K$, and network structure $[A_{nm}]$, a sufficiently large collection of control gains $F_n$ results in the formation of a stable fixed point at $z_n=z_n^*$. Our goal is to identify which oscillators require control (for which we will set $F_n>0$), and which oscillators do not require control (for which we will set $F_n=0$). To determine the control required we will first identify the target state $z_n^*$, then identify which oscillators require control.

\subsection{Target states}\label{subsec2_1}
We begin by finding the target states $z_n^*=\rho_n^*e^{i\theta_n^*}$ for each oscillator. To do so, we consider the equations of motion without control and search for a suitable steady-state. We assume that in the absence of control the system is not fully synchronized, an thus a stable fixed point of Eq.~(\ref{eq:LandauStuart}) does not exist. The target state will therefore represent the closest state to a fixed point which can be stabilized with control. 

Motivated by numerical exploration we present later, we will derive two different target states corresponding to two different control methods. For the first type, we assume that in addition to frequency synchronization, we wish to maintain a large amplitude for each oscillator, $\rho_n\approx 1$. In this case we simply the amplitude $\rho_n^*$ of each target state equal to one, and focus on finding an equilibrium of Eq.~(\ref{eq:Polar01}). For simplicity we linearize the sine term in Eq.~(\ref{eq:Polar01}), yielding an equilibrium characterized by
\begin{align}
0=\omega_n-K\sum_{m=1}^N L_{nm}\theta_m^*,\label{eq:Theta1}
\end{align}
or in vector form
\begin{align}
\bm{0}=\bm{\omega}-KL\bm{\theta^*},\label{eq:Theta2}
\end{align}
where $L$ is the network Laplacian matrix whose entries are defined $L_{nm}=\delta_{nm}k_{n}-A_{nm}$, where $k_n=\sum_{m}A_{nm}$ is the degree of node $n$. Although $L$ is singular (due to the fact that each row sums to zero), this equation can be solved using the Moore-Penrose pseudoinverse~\cite{Golub1996}. Specifically, given the eigenvalue decomposition $L=V^T\Lambda V$, whose eigenvalues can be ordered $0=\lambda_1<\lambda_2\le\dots\le\lambda_N$ such that $\Lambda=\text{diag}(\lambda_1,\lambda_2,\dots,\lambda_N)$ and the columns of $V$ are given by the eigenvectors of $L$, the pseudoinverse of $L$ is given by $L^\dagger=V^T\Lambda^{\dagger}V$, where $\Lambda^\dagger=\text{diag}(0,\lambda_2^{-1},\dots,\lambda_N^{-1})$. Applying the pseudoinverse to Eq.~(\ref{eq:Theta2}), we obtain
\begin{align}
\bm{\theta^*}=K^{-1}L^\dagger\bm{\omega}.\label{eq:Theta3}
\end{align}
Combined with unit target amplitudes, we obtain the target states $z_n^*=e^{i\theta_n^*}$ for type I control.

For the second type of target state, we relax the goal of driving oscillators to a large amplitude and therefore aim to find an approximate equilibrium of both Eqs.~(\ref{eq:Polar01}) and (\ref{eq:Polar02}). We begin by assuming that a given set of steady-state amplitudes $\rho_n^*$ are given, in which case an equilibrium of Eq.~(\ref{eq:Polar01}) after linearizing the sine term is characterized by
\begin{align}
0=\omega_n-K\sum_{m=1}^N\widehat{L}_{nm}(\bm{\rho^*})\theta_m^*,\label{eq:Theta4}
\end{align}
where $\widehat{L}(\bm{\rho^*})$ is the Laplacian matrix corresponding to the adjacency matrix $\widehat{A}(\bm{\rho^*})=P^{-1}AP$, where $P=\text{diag}(\rho_1^*,\dots,\rho_N^*)$. In vector notation Eq.~(\ref{eq:Theta4}) can be rewritten
\begin{align}
\bm{0}=\bm{\omega}-K\widehat{L}(\bm{\rho^*})\bm{\theta^*},\label{eq:Theta5}
\end{align}
Equation~(\ref{eq:Theta5}) can be solved similarly as Eq.~(\ref{eq:Theta2}) by applying the pseudoinverse, however, since $\widehat{L}(\bm{\rho^*})$ is not necessarily symmetric, its pseudoinverse is defined by the singular value decomposition. Specifically, if $\widehat{L}(\bm{\rho^*})=U^T\Sigma V$, where $\Sigma=\text{diag}(\sigma_1,\dots,\sigma_N)$ is populated by the real, nonnegative singular values $0=\sigma_1<\sigma_2\le\dots\le\sigma_N$ and the columns of $U$ and $V$ are populated by the left- and right-singular vectors, the pseudoinverse of $\widehat{L}(\bm{\rho^*})$ is defined $\widehat{L}^\dagger(\bm{\rho^*})=V^T\Sigma^\dagger U$, where $\Sigma^\dagger=\text{diag}(0,\sigma_2^{-1},\dots,\sigma_N^{-1})$. Applying the psuedoinverse to Eq.~(\ref{eq:Theta5}), we obtain
\begin{align}
\bm{\theta^*}=K^{-1}\widehat{L}^\dagger(\bm{\rho^*})\bm{\omega}.\label{eq:Theta6}
\end{align}

Shifting our attention now to the amplitudes, after expanding the cosine term to quadratic order, an equilibrium of Eq.~(\ref{eq:Polar02}) satisfies
\begin{align}
0&=\rho_n^*\left(1-\rho_n^{*2}\right)\nonumber\\
&+K\sum_{m=1}^NA_{nm}\left\{\rho_m^*\left[1-\frac{(\theta_m^*-\theta_n^*)^2}{2}\right]-\rho_n^*\right\},\label{eq:Rho1}
\end{align}
which is nonlinear and therefore cannot in general be solved analytically. However a solution can be obtained numerically given a collection of steady-stat phases $\theta_n^*$ using Newton's method. We note that Eq.~(\ref{eq:Rho1}) must be solve consistently with Eq.~(\ref{eq:Theta6}), which can be done iteratively. Specifically, we initialize $\bm{\theta^*}=\bm{0}$ and $\bm{\rho^*}=\bm{1}$ proceed iteratively. First, we obtain the next set of phases using Eq.~(\ref{eq:Theta6}), then solve Eq.~(\ref{eq:Rho1}) using Newton's method. Repeating this process, we converge onto our target state defined by $z_n^*=\rho_n^*e^{i\theta_n}$. We note that, in practice, when applied to Eq.~(\ref{eq:Rho1}) Newton's method can result in some unrealistic $\rho$ values, specifically $\rho_n^*>1$ or $\rho_n^*<0$. In such cases, we simply impose maximum and minimum values of $\rho_n^*=1$ or $\rho_n^*=\epsilon_\rho>0$, respectively. Together, the collection of $\theta_n^*$ and $\rho_n^*$ defined the target state $z_n^*=\rho_n^*e^{i\theta_n^*}$ for type II control.

\subsection{Control identification}\label{subsec2_2}
We now proceed to the question of identifying the oscillators that require control, assuming that target states $z_n^*$ have been computed as described above, either in the type I or II case. Assuming that the target state represents a fixed point of Eqs.~(\ref{eq:Polar01}) and (\ref{eq:Polar02}), its stability is indicated by the spectrum of the Jacobian matrix for the system. Specifically, the fixed point is stable if the real-part of the eigenvalues are contained in the left-half complex plane. Since our goal of consensus coincides with the stability of this given point, we aim to bound each eigenvalue to the left-half complex plane. Focusing on frequency-synchronization, we inspect Eq.~(\ref{eq:Polar01}), whose Jacobian matrix $DF$ is given by
\begin{align}
DF_{nm}=\left\{\begin{array}{rl}-K\sum \limits_{j\ne n}\widehat{A}_{nj}(\bm{\rho^*})\cos(\theta_j^*-\theta_n^*) &\text{if }m=n\\
 K\widehat{A}_{nm}(\bm{\rho^*})\cos(\theta_m^*-\theta_n^*)& \text{if }m\ne n\end{array}\right.\label{eq:DF1}
\end{align}
Importantly, the rows of $DF$ sum to zero -- a property that can be leveraged to identify any eigenvalues that may have positive real part and therefore destabilize the target state. In particular, after evaluating the Jacobian at the target state $z_n^*=\rho_n^*e^{i\theta_n^*}$, we define for each $n$ a radius $R_n=\sum_{m\ne n}|DF_{nm}|$ and a disc $D_n$ as the closed disc of radius $R_i$ centered at $C_n=DF_{nn}$. The Gershgorin circles theorem~\cite{Golub1996} ensures that all the eigenvalues of $DF$ lie within the union of all the Gershgorin discs. Specifically, since the rows of $DF$ sum to zero, if each off-diagonal entry of a row $n$ is positive, then $R_n=-DF_{nn}$ and it follows that the $n^{\text{th}}$ Gershgorin disc is contained in the left-half complex plane. However, if any off-diagonal entry of a row $n$ is negative, then $R_n>-DF_{nn}$, allowing the $n^{\text{th}}$ Gershgorin disc to partially enter the right-half complex plane, yielding the possibility of an eigenvalue with positive real part and a potential destabilization of the target state. 

By inspecting the Jacobian $DF$ evaluated at the target state, and specifically which rows contain negative off-diagonal entries, we can identify precisely which oscillators require control. In particular, with the addition of control the Jacobian becomes
\begin{align}
DF_{nm}=\left\{\begin{array}{rl}-K\sum \limits_{j\ne n}\widehat{A}_{nj}(\bm{\rho^*})\cos(\theta_j^*-\theta_n^*)-F_n&\text{if }m=n\\
 K\widehat{A}_{nm}(\bm{\rho^*})\cos(\theta_m^*-\theta_n^*)&\text{if }m\ne n\end{array}\right.\label{eq:DF2}
\end{align}
Specifically, for each row $n$ with negative off-diagonal entries, the control gain can be set to $F_n>R_n+DF_{nn}$, shifting the $n^{\text{th}}$ Gershgorin disc into the left-half complex plane, stabilizing the network.

Before proceeding to numerical examples, we make an important remark on the identification of oscillators that require control. In particular, the target phases and amplitudes $\theta_n^*$ and $\rho_n^*$ represent an approximation to the fixed point given the expansion of the sine and cosine terms in Eqs.~(\ref{eq:Polar01}) and (\ref{eq:Polar02}). In practice we build in a margin of error when identifying oscillators for control in order to overcome any inaccuracies induced by these aproximations. Specifically, rather than searching for rows with negative entries of the Jacobian, we set a threshold $\epsilon_\theta>0$ and identify any oscillator $n$ as requiring control if for any neighboring oscillator $m$ the entry $DF_{nm}/K\le \epsilon_\theta$.

\section{Numerical Examples}\label{sec3}

\begin{figure}[t]
\centering
\includegraphics[width=\linewidth]{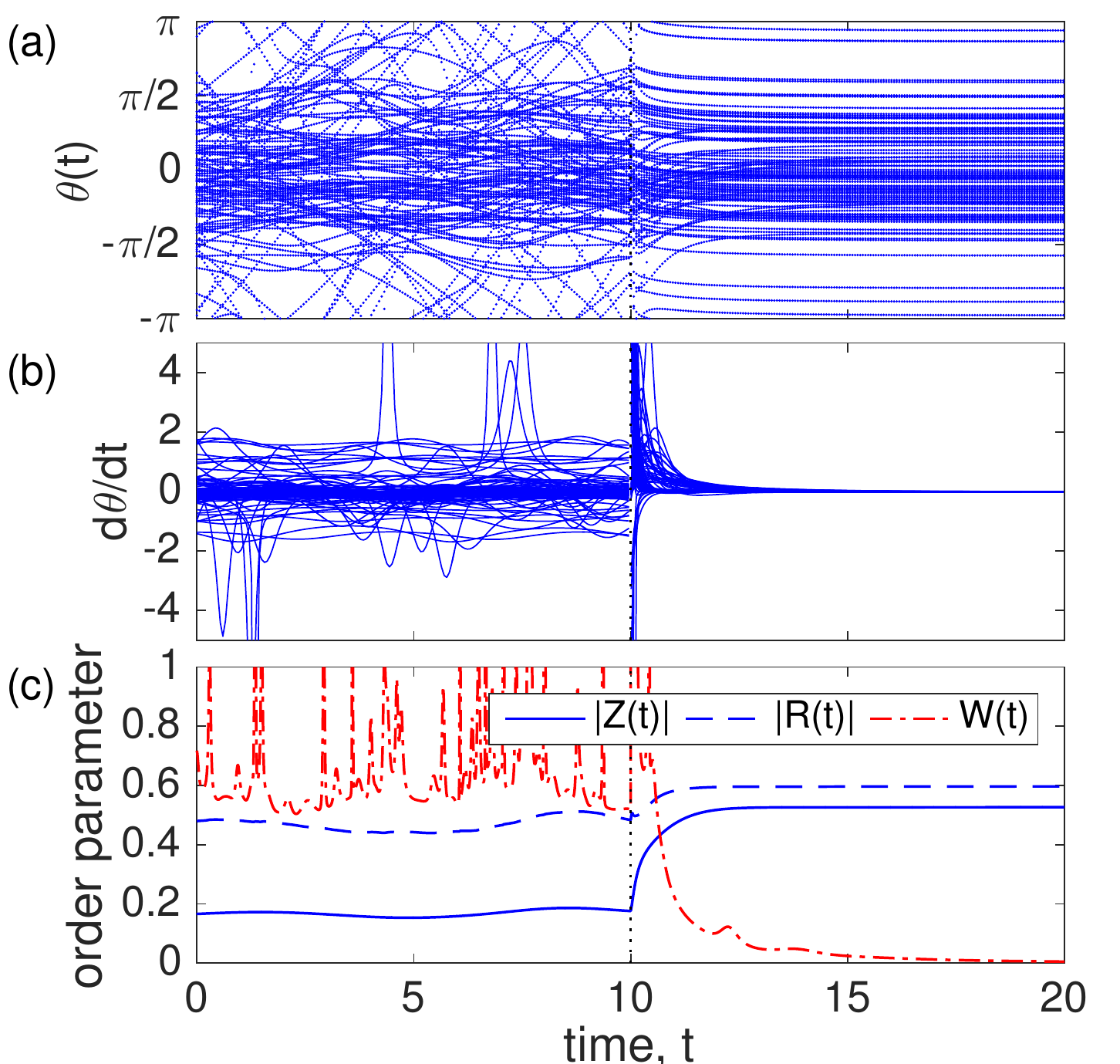}
\caption{(Color online) Type I control applied to an ER network of size $N=1000$ with mean degree $\langle k\rangle=6$, and coupling strength $K=0.3$. Time series of $10\%$ of the phases $\theta_n(t)$ and their angular velocities $t\theta_n/dt$ are plotted in panels (a) and (b), and the order parameters $|Z(t)|$ (solid curve) and $|R(t)|$ (dashed curve) are plotted in panel (c). Control is turned on at $t=10$. Natural frequencies are drawn from a uniform distribution of unit variance.}\label{fig1}
\end{figure}

\begin{figure}[t]
\centering
\includegraphics[width=\linewidth]{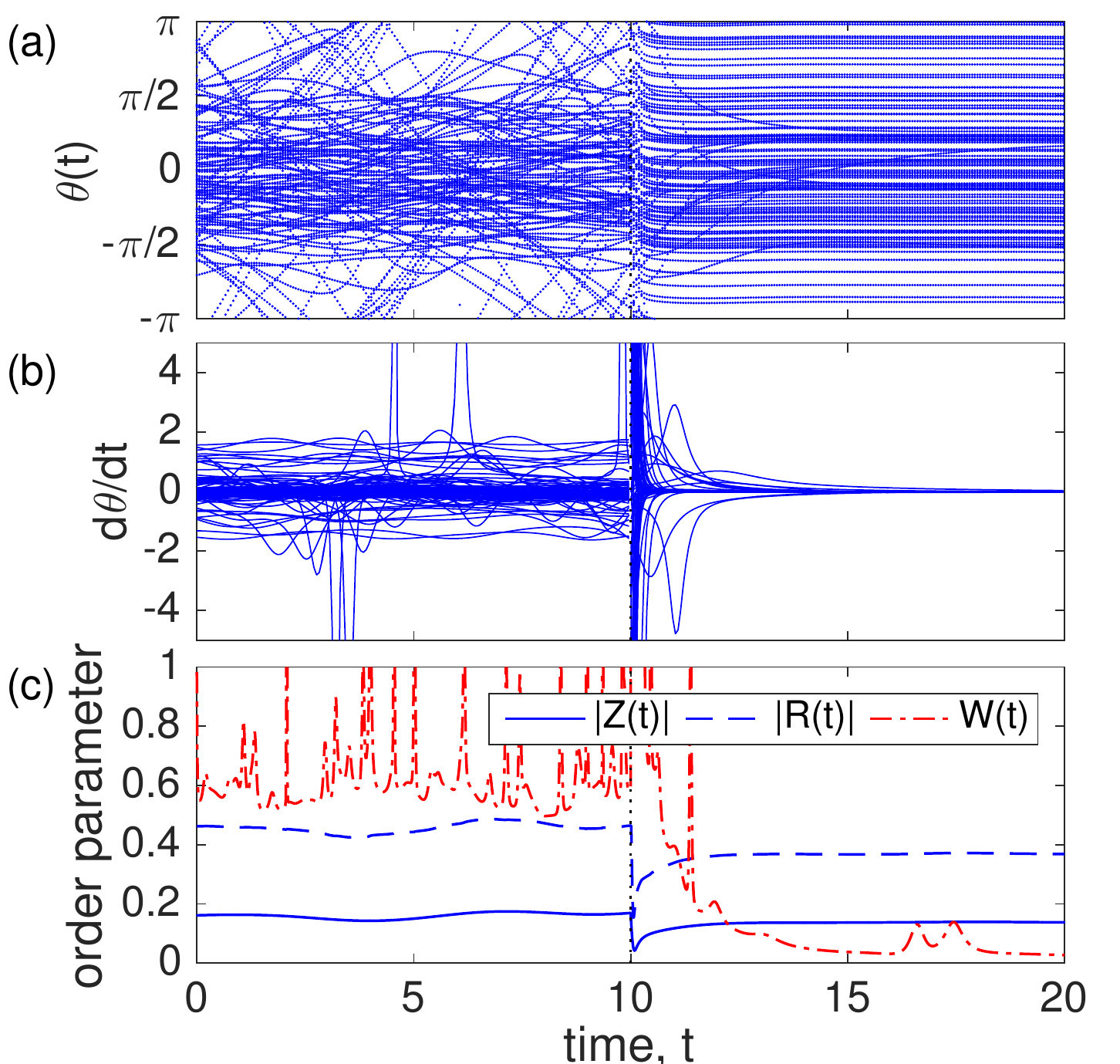}
\caption{(Color online) Type II control applied to the same network as in Fig.~\ref{fig1} ($N=1000$, $\langle k\rangle=6$, $K=0.3$). Time series of $10\%$ of the phases $\theta_n(t)$ and their angular velocities $t\theta_n/dt$ are plotted in panels (a) and (b), and the order parameters $|Z(t)|$ (solid curve) and $|R(t)|$ (dashed curve) are plotted in panel (c). Control is turned on at $t=10$.}\label{fig2}
\end{figure}

We now demonstrate the control method with several examples on random networks. Specifically, we use networks built using the Erd\H{o}s-Reny\'{i} (ER) model~\cite{Erdos1960}, where connections are created in a network of $N$ nodes in such a way that, for any given pair of nodes $(i,j)$, a link is made between nodes $i$ and $j$ with tunable probability $p$, resulting in a mean degree of $\langle k\rangle=p(N-1)$. Furthermore, in our simulations we consider oscillators whose natural frequencies are uniformly distributed with unit variance, i.e., drawn from the interval $[-\sqrt{3},\sqrt{3}]$. In the application of control we use threshold values (described above) of $\epsilon_\rho=0.2$ and $\epsilon_\theta=0.2$.

We begin by comparing type I and type II control implemented on an ER network of size $N$ with mean degree $\langle6\rangle$ and set the coupling strength to $K=0.3$. We plot the results of type I and type II control, respectively, in Figs.~\ref{fig1} and \ref{fig2}, plotting the time series of $10\%$ of the phases $\theta_n(t)$ in panels (a) and the corresponding angular velocities $d\theta_n/dt$ in panels (b). The plotted results are obtained by discarding a large transient and show the dynamics without control ($0\le t<10$) and after the control is turned on ($10\le t\le 20$). The vertical dotted lines at $t=10$ indicate control being turned on. For both type I and type II control the phases which are initially incoherent relax to equilibrium after the control is turned on, which can also be seen as the angular velocities relax to zero. 

\begin{figure}[t]
\centering
\includegraphics[width=\linewidth]{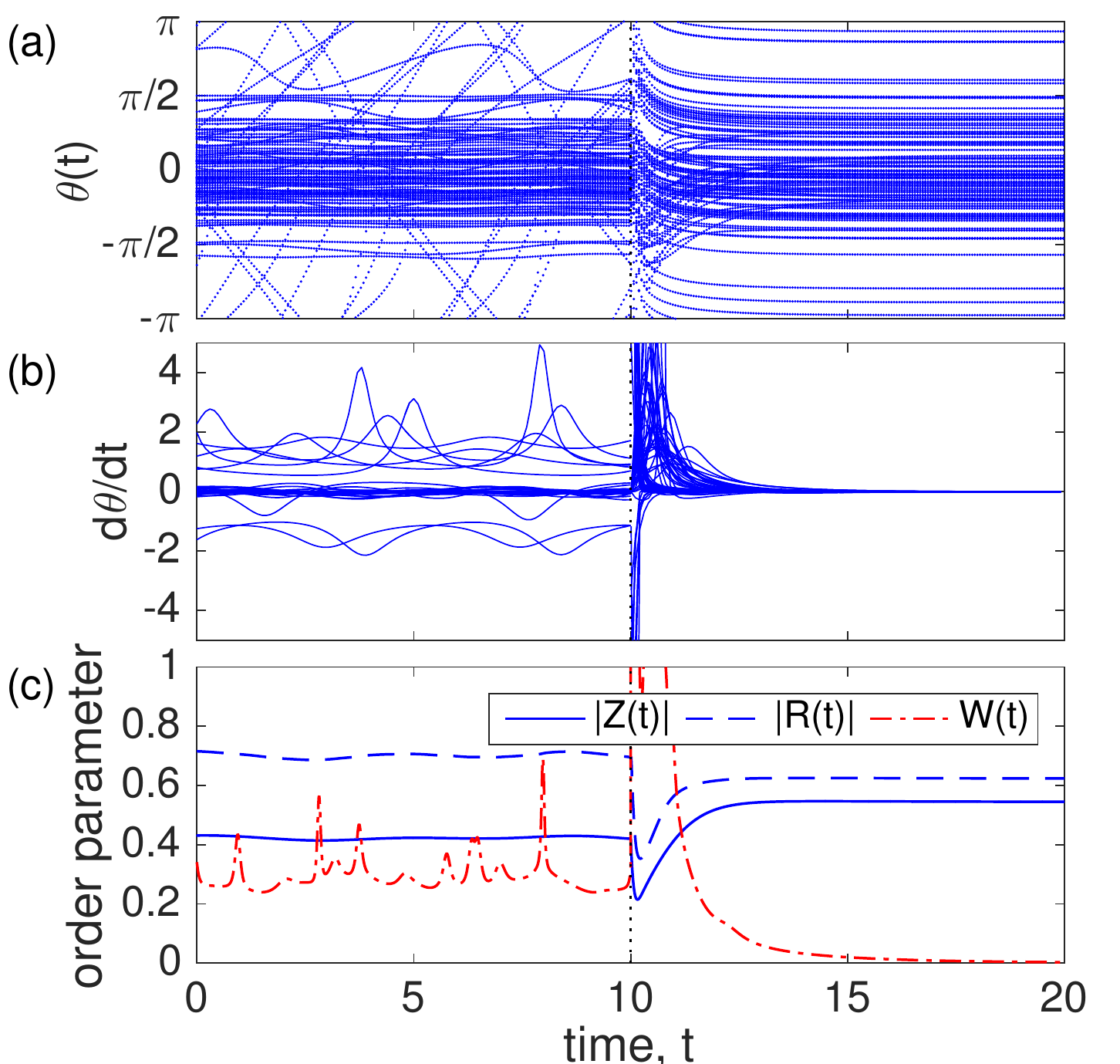}
\caption{(Color online) Type I control applied to the same network as in Figs.~\ref{fig1} and \ref{fig2} ($N=1000$, $\langle k\rangle=6$) but larger coupling strength, $K=0.4$. Time series of $10\%$ of the phases $\theta_n(t)$ and their angular velocities $t\theta_n/dt$ are plotted in panels (a) and (b), and the order parameters $|Z(t)|$ (solid curve) and $|R(t)|$ (dashed curve) are plotted in panel (c). Control is turned on at $t=10$.}\label{fig3}
\end{figure}

In addition to the time series of the phases and their angular velocities, we also consider three macroscopic order parameters. The first represents the mean field of the limit cycle oscillators:
\begin{align}
Z=\frac{1}{N}\sum_{n=1}^Nz_n=\frac{1}{N}\sum_{n=1}^N\rho_ne^{i\theta_n}.\label{eq:Ord01}
\end{align}
We also consider the classical Kuramoto order parameter which ignores the oscillators' amplitudes:
\begin{align}
R=\frac{1}{N}\sum_{n=1}^Ne^{i\theta_n}.\label{eq:Ord02}
\end{align}
The magnitudes of these complex order parameters, $|Z|$ and $|R|$, thus give meaningful measures of the network synchronization. Finally, as a complement to $Z$ and $R$, we consider an order parameter designed to quantify the instantaneous frequency dispersion in the network~\cite{Buzna2009PRE}:
\begin{align}
W=\sqrt{\frac{1}{N}\sum_{n=1}^N\left(\dot{\theta}_n-\left\langle\dot{\theta}\right\rangle\right)^2},\label{eq:Ord03}
\end{align}
where $\langle\cdot\rangle$ represents the mean over the population. We note that, while strong synchronization is typically indicated by larger values of $|Z|$ and $|R|$, small values of $W$ indicate small frequency dispersion, and therefore strong synchronization. We also plot in panels (c) of Figs.~\ref{fig1} and \ref{fig2} the time series of order parameters $|Z(t)|$ (solid blue curve), $|R(t)|$ (dashed blue curve), and $W(t)$ (dot-dashed red curve). As is typically the case (but does not need to be so), the Kuramoto order parameter $|R(t)|$ is larger than the order parameter $|Z(t)|$ due to the fact that $\rho_n\le1$ for all $n$. Interestingly, type I control enhances the degree of synchronization, as measured by both $|Z(t)|$ and $|R(t)|$, however type II control actually results in a smaller value of both. This can be explained by inspecting the distributions of target phases to which the oscillators relax in panels (a). In particular, type I control results in a distribution of phases that are clustered relatively close to the mean angle (here shifted to zero), while type II results in a much more uniform distribution of phases around the whole unit circle, resulting in a surprisingly low degree of phase synchronization. We also observe a large (and noisy) degree of frequency dispersion without control, however both type I and type II control drives the frequency dispersion $W(t)$ to a very small value.

\begin{figure}[t]
\centering
\includegraphics[width=\linewidth]{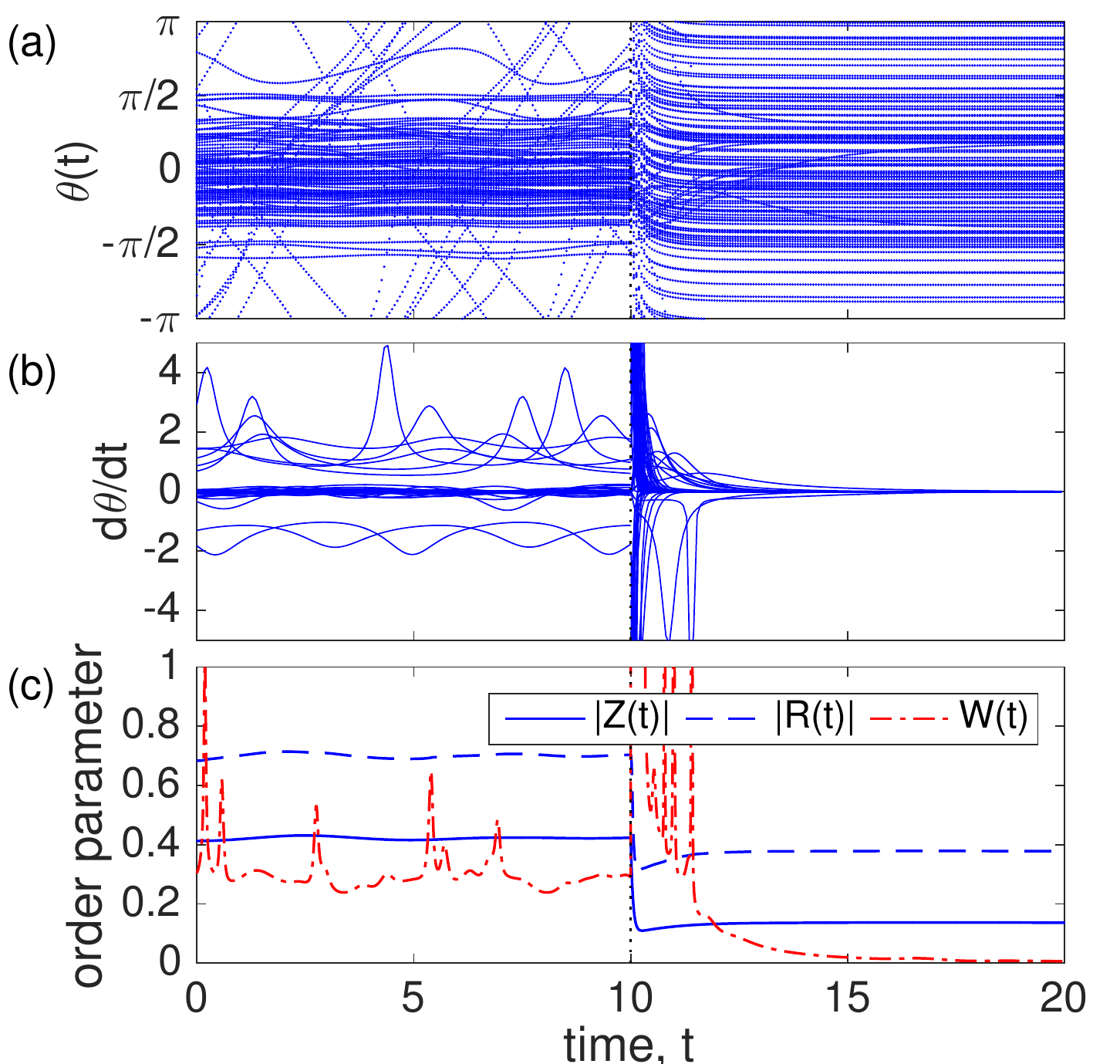}
\caption{(Color online) Type II control applied to the same network as in Fig.~\ref{fig3} ($N=1000$, $\langle k\rangle=6$, $K=0.4$). Time series of $10\%$ of the phases $\theta_n(t)$ and their angular velocities $t\theta_n/dt$ are plotted in panels (a) and (b), and the order parameters $|Z(t)|$ (solid curve) and $|R(t)|$ (dashed curve) are plotted in panel (c). Control is turned on at $t=10$.}\label{fig4}
\end{figure}

In the examples presented in Figs.~\ref{fig1} and \ref{fig2} the degree of synchronization of the initial state before control was relatively low, i.e., a significant number of oscillators were incoherent. We now contrast these experiments by considering the result of applying control to a state with initiall larger degree of synchronization. For the same network as used above, we increase the coupling strength to $K=4$, and present the results of applying both type I and type II control in Figs.~\ref{fig3} and \ref{fig4}. Note that before application of control the network is significantly more synchronized than for the case of $K=0.3$. In both cases of type I and type II control the network is able to relax to an equilibrium, however we note an interesting phenomenon with the macroscopic order parameters. In the case of type I control (Fig.~\ref{fig3}), while the order parameter $|Z(t)|$ increases, the Kuramoto order parameter $|R(t)|$ in fact decreases. More surprisingly, in the case of type II control (Fig.~\ref{fig4}), both order parameters $|Z(t)|$ decrease after control in applied. We find that this phenomenon can be attributed the the distribution of phases attained after control is applied. In particular, in both type I and type II control the distribution of steady-state phases can be relatively wide. This is more apparent in the case of type II control, but is also somewhat true for type I control. [Note that several oscillators in Fig.~\ref{fig3}(a) relax near $\theta\approx\pm\pi$.] Thus, while the control method can be used to attain consensus in the sense of frequency synchronization, the resulting state may be poorly phase-synchronized. This effect is curbed primarily in type I control for the order parameter $Z(t)$ since oscillators are explicitly driven to a larger amplitude, tending to result in an increased $|Z(t)|$. Finally, as in the example with smaller $K$, both type I and type II control drives the frequency dispersion $W(t)$ to a very small value.

\section{Failure of the method}\label{sec4}
Before closing, we briefly discuss the possible failures of the control method presented in this paper. While the method has been by-and-large effective in our explorations, we have also observed some cases where full frequency-synchronization is not attained. These failures deserve a few remarks. First, we note that in our explorations type I is very effective, failing very infrequently. Second, failures typically correspond to one or two oscillators remaining incoherent, and thus a very large fraction of the network end up in a state of frequency synchronization. Finally, we note that the likelihood of failure can be mitigated by modifying the threshold parameters $\epsilon_\rho$ and $\epsilon_\theta$. We emphasize that the target states $z_n^*=\rho_n^*e^{i\theta_n^*}$ which are central to the control method are approximations of theoretical equilibria of Eq.~\ref{eq:LandauStuart}, and thus it is to be expected that as this approximation fails, the likelihood of the control method failing increases. We find that the effect of these inaccuracies can be curbed by increasing these threshold parameters. We note that our explorations have focused on the case of ER random networks, and the presence of more complicated structural patterns in networks could affect the control method.

On the other hand, up to the approximations discussed above, the method presented here guarantees a spectrum of stable eigenvalues by ensuring that the Gershgorin circles, which contain the eigenvalues, are contained in the negative real-half of the complex plane. We note, however, that one or more Gershgorin circles partially crossing the imaginary axis into the real-half of the complex plane does not guarantee an unstable eigenvalue, but simply admits the possibility. Thus, it is likely possible that in some cases the synchronized state can be stabilized with less control (e.g., smaller control gains $F_n$) than suggested here.

\section{Discussion}\label{sec5}
In this paper we have investigated the control of networks of limit-cycle oscillators towards full frequency synchronization, i.e., consensus. The proposed method represents an extension of a method for controlling networks of phase oscillators~\cite{Skardal2015SciAdv} -- now applied to the limit-cycle counterpart given in Eq.~(\ref{eq:LandauStuart}). In particular, the method is based on identifying a target fixed point for the network and stabilizing this target state via an appropriately-defined feedback-type control. We proposed two types of control: type I control uses only the phase dynamics of the system and drives oscillators to large amplitudes (i.e., $\rho_n=1$), while type II control uses both the phase and amplitude dynamics of each oscillator to find a target state.

We have demonstrated the application of both type I and type II control with numerical examples and studied the effect that each type has on the macroscopic order parameters of the network. Surprisingly, the application of control can, and often does, decrease the degree of phase synchronization, as measure by the typical Kuramoto order parameter or it limit-cycle oscillator counterpart, while attaining strong frequency synchronization. We find that this phenomenon is due to the fact that, in many cases, the target state to which control drives the oscillators is relatively widely spread around the unit circle. Finally, we have included a discussion on the possible failures of the method. As the determination of the target state is based on an approximation, it is expected that as the approximation fails, the likelihood of the control method failing increases. We note, however, that this effect can be curbed by modifying threshold parameters built into the control identification process, and in cases of failure we find that only one or two oscillators remain desynchronized from the synchronized population.

We believe that the control method discussed in this paper complements other commonly used control methods such as pinning~\cite{Wang2002PhysA,Sorrentino2007PRE} and time-delay feedback~\cite{Choe2010PRE,Schneider2013Phil} which have in some cases been applied to limit-cycle oscillator dynamics such as those considered here, as well as other types of dynamics. While the work presented here can can be thought of as a type of feedback control method, we emphasize its novelty and simplicity -- stemming from a nonlinear dynamics stability analysis of the target synchronized state. More broadly, we believe that this work will be more generally useful for the control of nonlinear dynamics on complex networks and serve as inspiration for the development of control methods that combine essential elements from both the nonlinear dynamics of each unit and the structural properties of the network itself.

\acknowledgments
AA acknowledges support by the European Commission FET-Proactive project MULTIPLEX (Grant No. 317532), the ICREA Academia, the James S. McDonnell Foundation grant No. 220020325, and by FIS2015-71582.

\bibliographystyle{plain}

\end{document}